\documentclass[preprint,showpacs,groupedaddress,amsmath,amssymb]{revtex4}
\usepackage{graphicx}
\usepackage{dcolumn}
\usepackage{multirow}
\usepackage{amsfonts}

\newcommand{\beq}{\begin{equation}}
\newcommand{\eeq}{\end{equation}}
\newcommand{\bea}{\begin{align}}
\newcommand{\eea}{\end{align}}
\newcommand{\epsi}{\varepsilon}
\newcommand{\Cth}{\cos \theta}
\newcommand{\Sth}{\sin \theta}
\newcommand{\del}{\nabla}
\newcommand{\cS}{\mathcal{S}}
\newcommand{\PI}{\overleftrightarrow{\mathbf{\Pi}}}
\newcommand{\vphi}{\varphi}
\newcommand{\vtheta}{\vartheta}
\newcommand{\rhat}{\hat{r}}
\newcommand{\thetahat}{\hat{\vtheta}}
\newcommand{\phihat}{\hat{\vphi}}
\newcommand{\half}{\frac{1}{2}}
\newcommand{\wt}[1]{\widetilde{#1}}

\makeatletter
\def\Tbar{\mathchoice
   {\TTbar\displaystyle\textstyle{-}}%
   {\TTbar\textstyle\scriptstyle{-}}%
   {\TTbar\scriptstyle\scriptscriptstyle{-}}%
   {\TTbar\scriptscriptstyle\scriptscriptstyle{-}}%
   \!T}
\def\TTbar#1#2#3{{\setbox0=\hbox{$#1{#2#3}{\mathrm{T}}$}
     \raise2\p@\vbox{\hbox{$#2#3$}}\kern-.35\wd0}}
\makeatother

\begin{document}

\title{Non-Newtonian viscosity in magnetized plasma}


\author{Robert W. Johnson}
\email[]{rob.johnson@gatech.edu}
\affiliation{Fusion Research Center, Georgia Institute of Technology, Neely Building, Atlanta, GA 30332, USA}


\date{March 5, 2007.  Revised: September 26, 2007.}

\begin{abstract}
The particle and momentum balance equations can be solved on concentric circular flux surfaces to determine the effective viscous drag present in a magnetized tokamak plasma in the low aspect ratio limit.  An analysis is developed utilizing the first-order Fourier expansion of the poloidal variation of quantities on the flux surface akin to that by Stacey and Sigmar [Phys. Fluids, {\bf 28}, 9 (1985)].  Expressions to determine the poloidal variations of density, poloidal velocity, toroidal velocity, radial electric field, poloidal electric field, and other radial profiles are presented in a multi-species setting.  Using as input experimental data for the flux surface averaged profiles of density, temperature, toroidal current, toroidal momentum injection, and the poloidal and toroidal rotations of at least one species of ion, one may solve the equations numerically for the remaining profiles.  The resultant effective viscosities are compared to those predicted by Stacey and Sigmar and Shaing, {\it et al.}, [Nuclear Fusion, {\bf 25}, 4 (1985)].  A velocity dependence of the effective viscosity is observed, indicating the presence of non-Newtonian fluid behavior in magnetized plasma.
\end{abstract}


\maketitle


\section{Introduction}
The interpretation of measurements of rotational velocities in tokamak plasmas is an outstanding problem.  (See, for example, References~\cite{frc-pop-2006,solomonetal-pop-2006} and references therein.)  Since the pioneering work of Braginskii~\cite{brag-1965}, various authors~\cite{shaingetal-1985,stacandsig-1985,kim-soln-1991,nclasseqns-1997} have used a Newtonian fluid viscosity (which is not velocity dependent) to describe and predict the poloidal and toroidal rotational velocities in a magnetized plasma.  In this article, we expand upon the previous method~\cite{frc-pop-2006}, correcting the expression for the poloidal electric field along the way, to extract the experimentally observed poloidal and toroidal viscosities, which we then compare to neoclassical predictions~\cite{shaingetal-1985,stacandsig-1985}.  The effective viscosities are found to have a velocity dependence.

We start from the equations for particle and (angular) momentum balance (more properly termed force density or torque density balance) on concentric circular flux surfaces in the low aspect ratio limit $\epsi \equiv r/R_0 \ll 1$, where $R=R_0+r\Cth$, $\mathbf{B}=\mathbf{B}^0/(1+\epsi\Cth)=(0,B_\vtheta,B_\vphi)$, and $r \equiv \sqrt{\mathrm{Flux}_\vphi/(\pi B_\vphi^\mathrm{axis})}$, at steady-state, $\forall$ species~$j$:
\beq \label{eqn:continuity}
\del \cdot n_j \mathbf{V}_j + \frac{\partial n_j}{\partial t} \Rightarrow \del \cdot n_j \mathbf{V}_j = \cS_j \; ,
\eeq
where $\cS_j$ is the particle source for species~$j$, and
\beq \label{eqn:mombal}
m_j [ \del \cdot n_j \mathbf{V}_j \mathbf{V}_j ] + \del \cdot \PI_j + \del P_j - n_j e_j \left( \mathbf{E} + \mathbf{V}_j \times \mathbf{B} \right) - \mathbf{F}_j - \mathbf{M}_j = 0 \; .
\eeq
We work in SI units, including $kT$ in Joules, dropping Boltzmann's $k$---when temperatures are expressed numerically, the familiar conversion to eV will be applied.  (We suggest the use of the symbol $\Tbar$, much like the use of $\hbar$ in quantum physics, to represent a temperature that has absorbed Boltzmann's $k$.)  The pressure for species~$j$ is defined as $P_j = n_j \Tbar_j$, where $2 \Tbar_j \equiv m_j v_{\mathrm{th} \, j}^2$ and is assumed to have no poloidal dependence, {\it ie} $\partial \Tbar_j / \partial \theta = 0$.  The friction term $\mathbf{F}_j$ is taken in the form
\beq
\mathbf{F}_j = - \sum_{k \neq j} n_j m_j \nu_{jk} \left( \mathbf{V}_j - \mathbf{V}_k \right) \; ,
\eeq
where $\nu_{jk}$ is given by the NRL Formulary~\cite{physics-nrl}.  Our inertial term is in conservative form, $\PI_j$ is Braginskii's~\cite{brag-1965} rate-of-strain tensor, and $\mathbf{M}_j$ is the momentum (force) deposition density to species~$j$.  (We use $\mathbf{M}_j = n_j z_j^2 \mathbf{M} / \left< n z^2 \right>_j $, where $\left<\cdots\right>_j$ indicates the average over species, and assume $\mathbf{M}=(0,0,M_\vphi)$ for simplicity~\cite{stacey-words}.)  Various quantities are expanded on the flux surface in a first-order Fourier series; {\it eg} the density is expanded as
\beq \label{eqn:expan}
n_j (\theta) = n_j^0 \left( 1 + \epsi\wt{n}_j^c \Cth + \epsi\wt{n}_j^s \Sth \right) \; ,
\eeq
where $\epsi\wt{n}_j^{c/s} = n_j^{c/s}$ has unit normalization, and the flux surface averaged density is
\beq \label{eqn:fsavn}
\left< n_j \right> \equiv \frac{1}{2\pi} \oint_0^{2\pi} d\theta \left( 1+\epsi\Cth \right) n_j (\theta) = n_j^0 \left( 1 + \half \epsi^2 \wt{n}_j^c \right) \; .
\eeq
We also define the $\cos$ and $\sin$ flux surface moments as
\begin{align}
\left< n_j \right>_C & \equiv  \frac{1}{2\pi} \oint_0^{2\pi} d\theta \Cth \left( 1+\epsi\Cth \right) n_j (\theta) = \half \epsi n_j^0 \left( \wt{n}_j^c + 1 \right) \; , \mathrm{and} & \\
\left< n_j \right>_S & \equiv  \frac{1}{2\pi} \oint_0^{2\pi} d\theta \Sth \left( 1+\epsi\Cth \right) n_j (\theta) = \half \epsi n_j^0 \wt{n}_j^s \; . & 
\end{align}
Equation~(\ref{eqn:fsavn}) brings up an interesting point---the flux surface average of our expansion is not quite just the constant $n_j^0$ but includes a small term proportional to $\wt{n}_j^c$ as well.  Previous models~\cite{frc-pop-2006} have simply set $n_j^0$ (and the other similiar quantities) equal to the experimentally measured value, and so shall we, noting that this geometrical issue also affects the interpretation of the experimental measurements made at chords with a particular poloidal bias.  The radial derivatives of the poloidal coefficients continue to be neglected in this model.  Our coordinate system is a right-handed $(\rhat,\thetahat,\phihat)$ concentric toroidal flux surface geometry aligned so that the plasma current $\mathbf{I}_p$ points along the positive $\phihat$ axis, making $B_\vtheta > 0$ always.  Neutral beam injection along $\phihat$ can have either sign, as can the toroidal magnetic field $B_\vphi$.  We assume toroidal symmetry, so that $\partial / \partial\phi \rightarrow 0$.

\section{Development}

\subsection{Toroidal Equations}
The toroidal component of Equation~(\ref{eqn:mombal}) may be written ($\forall j$) as
\begin{align} \label{eqn:tormombal}
 & R \sum_{k \neq j} n_j m_j \nu_{jk} \left( V_{\vphi \, j} - V_{\vphi \, k} \right) - \frac{1}{rR} \frac{\partial}{\partial r} \left[ R^3 \left( \frac{n_j \Tbar_j}{\nu_{\vphi \, j}^{drag}} \right) \frac{\partial}{\partial \theta} \frac{V_{\vphi \, j}}{R} \right] & \\
 & = R M_{\vphi \, j} + R n_j e_j \left( E_\vphi + B_\vtheta^0 V_{r \, j} \right) \equiv \cS_{\vphi \, j} \; , &
\end{align}
where now $\cS_{\vphi \, j}$ is the toroidal torque source term and the factor of $R$ is necessary because of $R$'s dependence on $\theta$.  That we may replace the gyroviscous term with an effective viscous term of the form above is shown by, for a general frequency $\nu_{\vphi \, j}^{drag}$ with no poloidal dependence,
\beq
\left< R \; \phihat \cdot \del \cdot \PI_{34}^{\nu_{\vphi \, j}^{drag}} \right> \simeq - \left< \frac{1}{rR} \frac{\partial}{\partial r} \left[ R^3 \left( \frac{n_j \Tbar_j}{\nu_{\vphi \, j}^{drag}} \right) \frac{\partial}{\partial \theta} \frac{V_{\vphi \, j}}{R} \right] \right> \equiv \left( \frac{n_j^0 \Tbar_j}{\nu_{\vphi \, j}^0} \right) \frac{V_{\vphi \, j}^0}{R_0} \; ,
\eeq
where we have absorbed some poloidal coefficients resulting from the integration into our parameter $\nu_{\vphi \, j}^0$.  Doing so facilitates isolating these equations from the rest.  (Note the correct placement of $\nu_{\vphi \, j}^0$ necessary to keep the units right.)  Braginskii's $\eta_j^4 \simeq \eta_j^0/\Omega_j\tau_j \simeq n_j \Tbar_j / \Omega_j$, for the gyrofrequency $\Omega_j$ and the species collision time $\tau_j \equiv 1/\nu_{jj}$.  Substituting for the gyrofrequency $\Omega_j = e_j B / m_j$ recovers the familiar expression $\eta_j^4 \simeq n_j \Tbar_j m_j / e_j B$.  Our evaluation of the gyroviscous contribution ({\it cf.} Equation~(12) in Reference~\cite{frc-pop-2006}) is
\begin{align} \label{eqn:gyrovisc}
1 / \nu_j^{gyro} \equiv & \frac{-R_0}{n_j^0 \Tbar_j V_{\vphi \, j}^0} \left< \frac{1}{rR} \frac{\partial}{\partial r} \left[ R^3 \left( \frac{n_j \Tbar_j m_j}{e_j B_\vphi} \right) \frac{\partial}{\partial \theta} \frac{V_{\vphi \, j}}{R} \right] \right> & \\
 = & - \half \frac{m_j}{e_j B_\vphi^0} \left( \wt{n}^s_j + 4 \wt{V}_{\vphi \, j}^s \right) & \\
 &  - \epsi \half \frac{m_j R_0}{e_j B_\vphi^0} \left[ \wt{n}^s_j \left(1-\wt{V}_{\vphi \, j}^c\right) + \wt{V}_{\vphi \, j}^s \left(4+\wt{n}^c_j\right) \right] \left( \frac{1}{V_{\vphi \, j}^0}\frac{\partial}{R_0 \partial \epsi}V_{\vphi \, j}^0 + \frac{1}{P_j^0}\frac{\partial}{R_0 \partial \epsi}P_j^0 \right) \; . &
\end{align}
Stacey and Sigmar~\cite{stacandsig-1985} apparently drop the leading order term, and so shall we when evaluating their prediction for the gyroviscosity, for consistency.  Its inclusion does not change the magnitude of the gyroviscosity profiles, as the terms may be seen to have the same units-containing coefficient upon substituting for the radial derivative in terms of $\epsi$, but does relocate the zeros.  Our effective viscous term incorporates effects from the gyroviscous and any ``anomalous'' terms that may be present.  The effective toroidal torque transfer frequency for species $j$ is identified as
\beq
1 / \nu_{\vphi \, j}^0 \equiv \left< 1 / \nu_{\vphi \, j}^{drag} \right> = \frac{-R_0}{n_j^0 \Tbar_j V_{\vphi \, j}^0} \left< \frac{1}{rR} \frac{\partial}{\partial r} \left[ R^3 \left( \frac{n_j \Tbar_j}{\nu_{\vphi \, j}^{drag}} \right) \frac{\partial}{\partial \theta} \frac{V_{\vphi \, j}}{R} \right] \right> \; ,
\eeq
giving an effective toroidal viscosity of $\eta^\vphi_j = n_j^0 \Tbar_j / \nu_{\vphi \, j}^0$.  If at least one species' toroidal velocity profile is known (as well as the source terms $\cS_{\vphi \, j}$), the Equations~(\ref{eqn:tormombal}) may be used to determine the remaining species' toroidal velocity profiles, as well as the species' averaged toroidal torque transfer frequency $\nu_\vphi^0 \equiv \nu_{\vphi \, j}^0 \left< P_j^0 \right>_j / P_j^0 $, by equating the species' effective toroidal viscosities.

\subsection{Radial Electric Field and Poloidal Velocity Profiles}
The radial component of Equation~(\ref{eqn:mombal}) yields the familiar expression for the radial electric field:
\beq \label{eqn:erad}
E_r = - \frac{\partial \Phi}{\partial r} = \frac{1}{n_j e_j} \frac{\partial P_j}{\partial r} + B_\vtheta V_{\vphi \, j} - B_\vphi V_{\vtheta \, j} \; .
\eeq
Knowledge of the density normalized pressure gradient and the poloidal and toroidal velocity profiles for one species is used to determine the (flux surface averaged) radial electric field.  The lack of a species index~$j$ on $E_r$ means that knowledge of the remaining species' pressure gradients and toroidal rotation profiles, as given in the subsection above, sufficiently determines the remaining species' poloidal velocity profiles.  Furthermore, we may take the $\cos$ and $\sin$ flux surface moments of Equation~(\ref{eqn:erad}) to gain expressions for the poloidal variation of the radial electric field,
\begin{align}
\wt{E}_r^c &= -1 + \left( \frac{1}{n_j^0 e_j} \frac{\partial P_j^0}{\partial r} + B_\vtheta^0 V_{\vphi \, j}^0 \wt{V}_{\vphi \, j}^c - B_\vphi^0 V_{\vtheta \, j}^0 \wt{V}_{\vtheta \, j}^c \right) / E_r^0 \; , \mathrm{and} & \\
\wt{E}_r^s &= \left( B_\vtheta^0 V_{\vphi \, j}^0 \wt{V}_{\vphi \, j}^s - B_\vphi^0 V_{\vtheta \, j}^0 \wt{V}_{\vtheta \, j}^s \right) / E_r^0 \; , &
\end{align}
valid $\forall j$.  These equations require the poloidal variations of the poloidal and toroidal velocities, which we come to next.

\subsection{Poloidal Equations}
\subsubsection{Poloidal Force Density Balance}
We write the poloidal component of Equation~(\ref{eqn:mombal}) as
\beq \label{eqn:polmombal}
m_j \thetahat \cdot \del \cdot n_j \mathbf{V}_j \mathbf{V}_j + \thetahat \cdot \del \cdot \PI_j + \frac{1}{r} \frac{\partial P_j}{\partial \theta} - n_j e_j \left( E_\vtheta - B_\vphi V_{r \, j} \right) = F_{\vtheta \, j} + M_{\vtheta \, j} \; ,
\eeq
where the inertial term is given by
\beq
m_j \thetahat \cdot \del \cdot n_j \mathbf{V}_j \mathbf{V}_j = m_j n_j \left[ V_{r \, j} \frac{\partial V_{\vtheta \, j}}{\partial r} + \frac{V_{\vtheta \, j}}{r} \left( V_{r \, j} + \frac{\partial V_{\vtheta \, j}}{\partial \theta} \right) +V_{\vphi \, j}^2 \frac{\Sth}{R} \right] + m_j  V_{\vtheta \, j} \del \cdot n_j \mathbf{V}_j \; ,
\eeq
and the viscous term~\cite{stacandsig-1985} is
\beq
\thetahat \cdot \del \cdot \PI_j = \frac{1}{r} \frac{\partial}{\partial \theta} \left( \eta_j^0 \frac{A_j^0}{2} \right) - \frac{3 \Sth}{R} \left( \eta_j^0 \frac{A_j^0}{2} \right) \; ,
\eeq
for
\beq
\half A_j^0 = - \frac{1}{3 r} \frac{\partial V_{\vtheta \, j}}{\partial \theta} + \frac{V_{\vtheta \, j}}{r} \left( \frac{1}{R} \frac{\partial R}{\partial \theta} + \frac{1}{3 B_\vtheta} \frac{\partial B_\vtheta}{\partial \theta} \right) + \frac{B_\vtheta R}{B_\vphi r} \frac{\partial V_{\vphi \, j}/R}{\partial \theta} \; .
\eeq
While a factor numerically equal to the safety factor $q \equiv | B_\vphi r / B_\vtheta R | $ is apparent, we eschew writing our equations in terms of $q$ because of the complications arising from keeping track of the sign.  Shaing and Stacey~\cite{shaingetal-1985} propose to modify Braginskii's parallel viscosity, $\eta_j^0 = n_j \Tbar_j \tau_j$, for the banana-plateau regime, $\nu^\star_j \equiv \nu_{jj} q R / v_{th \, j} \epsi^{3/2} \sim 1$, by accounting for trapped particle effects with a plateau-suppressed collision time which translates into our notation as
\beq \label{eqn:shaing}
\tau^\star_j \equiv \frac{2 \left( q R_0 / v_{th \, j} \right) \nu^\star_j}{\left(1 + \nu^\star_j \right) \left(1 + \epsi^{3/2} \nu^\star_j \right)} \; ,
\eeq
which we use to compare the predicted parallel viscosity $n_j \Tbar_j \tau^\star_j$ with our effective viscosity evaluated by treating $\tau^\star_j$ as a free parameter of the theory.

For each species~$j$, we now develop a system of 5 nonlinear equations in 5 unknowns, coupled through the friction terms:  the unity, $\cos$, and $\sin$ flux surface moments of Equation~(\ref{eqn:polmombal}), and the $\cos$ and $\sin$ flux surface moments of Equation~(\ref{eqn:tormombal}).  The poloidal variations in density are simply related to the poloidal variations of the poloidal velocity via continuity, Equation~(\ref{eqn:continuity}), as $\wt{V}_{\vtheta \, j}^c = - ( \wt{n_j}^c + 1 )$ and $\wt{V}_{\vtheta \, j}^s = - \wt{n_j}^s$, when radial flows are neglected.

\subsubsection{Poloidal Electric Field}
In order to evaluate Equation~(\ref{eqn:polmombal}), we need the proper expression for the poloidal electric field, one which does not depend on what must be a physically insignificant absolute potential as in previous theories~\cite{frc-pop-2006}.  (See pages 81-82 of Reference~\cite{griffiths-1989-81} for a clear discussion of the issue.)  Writing that equation for the electron species ($e_e = -e$), retaining only the gradients of pressure and electric potential, and inserting our poloidal expansion from Equation~(\ref{eqn:expan}) in terms of $n_e^{c/s}$ (no tilde) gives us
\beq
\frac{\Tbar_e}{e \, r} \left( n_e^s \Cth - n_e^c \Sth \right) = \left( 1 + n_e^c \Cth + n_e^s \Sth \right) \frac{1}{r} \frac{\partial \Phi}{\partial \theta} \; .
\eeq
Taking the flux surface average of both sides of the equation above and identifying terms which survive eventually yields the interesting result that the poloidal electric field must vanish
\beq \label{eqn:polE}
E_\vtheta \left( \theta \right) \equiv 0 \; ,
\eeq
to our order of approximation.  Any nonzero portions of $E_\vtheta$ lead to coordinate singularities in the theory which are unphysical.  The crux of the argument is that the integral must hold $\forall \; r$, $n_e^c$, and $n_e^s$, independently.

\subsubsection{Radial Electric Field Revisited}
A vanishing $E_\vtheta$ gives us a (D-shaped) handle on the radial electric field.  Consider the line integral of $\mathbf{E}$ around the closed path shown in Figure~1
 for an arbitrary flux surface at $r$.  Since $\oint \mathbf{E} \cdot d\mathbf{l} \equiv 0$, the integral along path A cancels that of path B.  That $\int_0^{\pi} E_\vtheta r d\theta = 0$ follows directly from Equation~(\ref{eqn:polE}), thus the integral along A vanishes as well.  That integral may be written as the sum of integrals along the negative and positive x-axis, which for our usual poloidal expansion (no tildes) yields
\begin{align}
0 & = \int_r^0 E_r(\pi) dr + \int_0^r E_r(0) dr & \\
 & = -\int_0^r ( E_r^0 - E_r^c ) dr + \int_0^r ( E_r^0 + E_r^c ) \; . &
\end{align}
Rotating the given path by $\pi/2$ recovers a similar expression for $E_r^s$.  Thus, the poloidal variations of the radial electric field must vanish (to our order of expansion at least if not to all).

\section{Input Profiles and Evaluation}
To solve our system of equations, six for each species~$j$ including the toroidal equations (given fully in Appendix~\ref{app:eqns}, suitably normalized), we need to input profiles for density, temperature, poloidal magnetic field (intrinsically related to the toroidal current, hence the toroidal electric field as well), toroidal momentum injection, and the rotation profiles for at least one species of ion.  To analyze shots from DIII-D~\cite{diiid-2002}, we retrieve the necessary magnetic geometry and density and temperature profiles from EFIT~\cite{Lao:1985mw}, the rotation profiles for Carbon-6 come from GAProfiles~\cite{gaprof-2000}, and the momentum injection is calculated by NBEAMS~\cite{nbeams-1992}.  [We make use of data provided by the Atomic Data and Atomic Structure (ADAS) database---the originating developer of ADAS is the JET Joint Undertaking.] The toroidal electric field is evaluated from the plasma current profile using Braginskii's~\cite{brag-1965} resistivity and Wesson's~\cite{tokamaks-2004} Coulomb logarithm.  We neglect the electron and radial flows in the following analysis---a more thorough treatment including such flows is forthcoming.

We follow the outline of the previous section to calculate first the effective toroidal viscosity profile and the toroidal rotation profiles for each species other than Carbon-6, and then those species' poloidal rotation profiles.  We also investigated modifying the toroidal equations to reflect equal torque transport frequencies rather than viscosities---the effective viscosities split, but not by much, and the deuterium's velocity drops significantly.  Intuitively, the viscosities are the physical quantities which should equate, not the frequencies, so we pursue equal effective toroidal viscosities in this article.  Armed with the velocity profiles, we enter the poloidal analysis to compute the radial profiles of the effective poloidal viscosity, of the poloidal (Fourier) components of each species' density, poloidal velocity, and toroidal viscosity, and the predictions of Shaing's~\cite{shaingetal-1985} parallel viscosity and Stacey's~\cite{stacandsig-1985} gyroviscosity.  The vanishing of the radial electric field's poloidal coefficients lets us eliminate two of our nonlinear force density equations in favor of those linear equations---we choose the toroidal $\cos$ and $\sin$ moments.  Reemploying those equations would let us solve for two more quantities, such as the neglected poloidal coefficients of temperature or the poloidal dependence of $\nu_{\vphi \, j}^{drag}$.  To solve these equations at each flux surface, we generally employ \textsc{Matlab}'s trust-region based solver~\cite{fsolv1-1999,fsolv2-1988} or Levenberg-Marquardt~\cite{lmalg-1977} algorithm. 

\section{Results for Shot 98777 @ 1600ms}
\subsection{Preliminaries}
Our most thoroughly investigated shot is the L-mode 98777.  As we currently are restricted to a steady state analysis, we select the time 1600ms for its relatively stable density.  The input profiles are displayed in Figure~2.
  The deuterium density is determined from the electron density and the carbon density via charge neutrality, $n_e = \sum_{ions} z_j n_j$.  We use the recently corrected~\cite{solomonetal-pop-2006} toroidal rotation profile and the uncorrected poloidal velocity profile---not necessarily the most consistent approach, but until corrections to the poloidal velocity for early shots are available, the best we can do for the shot at hand.  Our qualitative results would not be affected by a small shift in the carbon poloidal velocity, just the actual numbers.  Our calculated radial electric field is in Figure~3
---applying the species index to $E_r$ indicates the value from the RHS of Equation~(\ref{eqn:erad}).  Surprisingly (or not, for nonlinear equations), we find two good solutions, one that is friction dominated, with a lower viscosity, and one that is viscosity dominated, with a higher viscosity.  We stress that both solutions are compatible with the physics developed in this article, but obviously the experiment realized just one.  A further constraint from the experiment which must have been neglected (or unmeasured) would be needed to decide between the two.  An interesting method to resolve the issue might be to analyze a helium discharge with (corrected) measurements for both the helium and carbon velocities but treating the helium species as ``unknown'' for the purposes of the present theory.  As we can find no reasonable prejudice against either, we will discuss both solutions.

\subsection{Friction Dominated Solution}
The profiles for the poloidal coefficients of density and poloidal and toroidal velocity for the friction dominated solution are displayed in Figure~4.
In order to get both positive and negative values on a single logarithmic scale, negative values are depicted with an ``o'', and positive values are as in the legend.  We see that, generally,  the sine coefficients start at $\mathcal{O}(10^{-5})$ at the core and approach unity at the edge, while the cosine coefficients begin around $\mathcal{O}(10^{-2})$ and approach unity at the edge.  The profiles for the velocities and viscosities are in Figure~5.
The deuterium's toroidal velocity remains within a few percent of the carbon's value, and the effective toroidal viscosity remains right around $\mathcal{O}(10^{-5})$ for the entire radial profile with positive sign.  (One should note that there is a $-$ sign which may either be written explicitly in front of the drag term or absorbed into the effective viscosity.  We have done the latter.)  While the prediction for carbon's gyroviscosity is down another $\mathcal{O}(10^{-5})$, the prediction for deuterium's gyroviscosity is in excellent agreement with the experimentally observed viscosity resulting from the preceding analysis for the majority of the profile---given our assumption that the species' effective viscosities are equal.  We note that the zeros present in the $V_\vphi^s$ profiles make an appearance in the gyroviscosity profiles (they dominate the $n_j^s$ contribution to Equation~(\ref{eqn:gyrovisc}) by a factor of 4) and are associated with zeros of the relevant poloidal velocity profiles.

For the poloidal velocity, we find that again the deuterium's value is within some percent of the carbon's, albeit this time there is a greater difference, especially in the core of the plasma.  We also observe that, over much of the unit-normalized radius $\rho$, the effective viscosities are several orders of magnitude above Shaing's parallel viscosity, with some interesting interruptions by nonlinear behavior.  Where the viscosity spikes we interpret as the plasma ``going glassy'' at that $\rho$, and where the viscosity plummets we interpret as the plasma ``thinning out''.  Around $\rho = .64$, and again out near the edge around $\rho = .98$, the carbon viscosity appears to reach a glassy peak where the poloidal velocity goes through zero.  The deuterium viscosity exhibits the same behavior where its velocity crosses or approaches zero, as well as follows the general shape of Shaing's~\cite{shaingetal-1985} parallel viscosity upto a roughly constant factor.  A viscosity which depends on more than just the pressure is associated with the behavior of a non-Newtonian fluid, and higher viscosity at lower velocity indicates that a magnetized plasma behaves like a pseudoplastic fluid.  We also note that, interestingly, the gyroviscosity is predicted to thin where correspondingly the effective poloidal viscosity goes glassy, and that while the magnitudes are off, the smoothness of the predicted parallel viscosity is reflected in the effective toroidal viscosity---perhaps the underlying physics in the neoclassical derivations of those quantities ought to be readdressed, though more likely our equating the effective toroidal viscosities is masking the independent behavior of the species.

\subsection{Viscosity Dominated Solution}
For the viscosity dominated solution's poloidal coefficient profiles, Figure~6,
we see less variation between the species in the sine coefficients than in the friction dominated solution, about the same variation in the cosine coefficients for density and poloidal velocity, and a distinguishingly large variation between the species' cosine coefficients of the toroidal velocity, $V_\vphi^c$.  For this solution, the asymmetry coefficients of the toroidal velocities seem to be more related than previously, for now the carbon's sine profile is reflected in the deuterium's cosine profile as zeros appear in identical locations.  That deuterium's $V_\vphi^s$ exceeds unity at the edge we interpret as the toroidal rotation changing sign---if any density coefficients had exceeded unity (as prior models' did~\cite{frc-pop-2006}), that would signify a breakdown of our theory.  Luckily, they remain within bounds.  It would be interesting to learn if the behavior of $V_\vphi^s$ is associated with the presence of a divertor.

The velocity and viscosity profiles, showing this solution's unexpected predictions for the deuterium velocities, are in Figure~7.
The gyroviscosity for carbon again exhibits the zeros in the $V_\vphi^s$ profile and generally underpredicts the effective viscosity by a factor of $\mathcal{O}(10^{-5})$.  Deuterium's gyroviscosity is within a factor of 10 to 100 for the majority of the profile.  The effective toroidal viscosities now have the sign $-$.  The prediction for the poloidal velocity of deuterium does not seem unusual in light of the formula used for its evaluation, Equation~(\ref{eqn:erad}), as both species are constrained to have the experimentally determined radial electric field.  Again, the effective poloidal viscosity for carbon exhibits pseudoplastic behavior, peaking at values slightly above those of the deuterium, and settling to a value just under.  Other possible solutions were in good agreement with Shaing's parallel viscosity over much of the profile, with the same glassy behavior at zeros of the poloidal velocity.  Deuterium's effective viscosity again resembles the prediction for carbon up to a consistent factor.  For both the friction and the viscosity dominated solutions, the effective poloidal viscosity for each species approaches a common value at the core and at the edge.

\subsection{Interpretation}
We may recover the functional dependence of the poloidal viscosity on the poloidal velocity by plotting the log of (the absolute value of) both $\tau_j^{\star}$ versus $V_{\vtheta \, j}$, Figure~8.
We combine the results from both solutions above, as well as the results from a similar analysis of the H-mode shot 99411 at 1800ms, and separate the deuterium, Figure~8(a),
from the carbon, Figure~8(b).
Across eight orders of magnitude, the deterium points lie on a single straight line fit, which we label L1, with a slope $-S$ of -1.032(8) and an intercept $\sim 1.7$.  For carbon, line L2 lies a little lower, with a slope $-S$ of -1.019(9) and an intercept about $.1$ less than L1's.  Thus, we believe that $\tau_j^{\star} \propto (V_{\vtheta \, j})^{-S}$, noting that the statistical error indicates a value for $S$ slightly different than unity.  (There needs to be another velocity in the numerator to cancel the units---see below.)  The other possible solutions mentioned above return a plot for carbon that exhibits multiple lines at roughly equal slopes and equidistant intercepts, thus we suppose that the coefficient depends on the charge-state $z_j(\rho)$ to account for the quantization of carbon's lines.  That carbon may have behavior consistent with multiple charge-states coexisting is not currently accounted for in the model, which treats all carbon as having $z_C = 6$.

The unexpected behavior found for the effective viscosities has a simple interpretation. As the plasma velocity $V_{\alpha \, j}$ approaches from above some critical velocity $V_{\alpha \, j}^{crit}$, whose physics has yet to be determined, the ions of species~$j$ may encounter a greater number of collisions with ions of neighboring flux surfaces than at higher fluid speeds (with more collimated particle behavior---consider urban highway traffic during rush hour for an analogy, when a high density may flow at independently high speeds in different lanes until one vehicle slows down to exit, prompting drivers behind to switch to lanes to either side), thereby impeding their progress past each other in successive layers.  Or perhaps when ions slow down they become more impeded by the negatively charged electrons mostly neglected in this work.  Either way, the effective poloidal viscosity clearly displays a dependence on the poloidal velocity, while also in fair agreement with Shaing's~\cite{shaingetal-1985} parallel viscosity over much of the profile.  Such behavior could be accounted for by applying to the dynamic viscosity a simple correction factor $\Gamma$ of the form $\eta^\alpha_j \rightarrow \eta^\alpha_j \Gamma (|V_{\alpha \, j}^{crit}| / V_{\alpha \, j})$.  In order to avoid the divergence at $V_{\alpha \, j} = 0$, we suggest the factor $\arctan (|V_{\alpha \, j}^{crit}| / V_{\alpha \, j})$, with appropriate normalization.  That factor picks up the sign of $V_{\alpha \, j}$, asymptotes to zero on a scale set by $V_{\alpha \, j}^{crit}$, and approaches finite limits of opposite sign and equal magnitude as $V_{\alpha \, j}$ approaches zero from either side---precisely the behavior we observe in the effective viscosity.  The use of $\arctan$ suggests that a ratio of orthogonal vectors is involved in the physics, but another functional form which reflects what might be a significant difference from unity in the power on $V_{\alpha \, j}$ might be necessary.  Most likely $V_{\alpha \, j}^{crit}$ is on the scale of the radial velocity $V_{r \, j}$, but more detailed investigation of the underlying physics is needed for any certain determination.  That the toroidal gyroviscosity is predicted to thin at locations where the poloidal effective viscosity goes glassy is noted, as well as the absence of peaks in the parallel viscosity which matches the smooth profile found for the toroidal effective viscosity, suggesting the possibility that the physics assumed to apply toroidally might better be applied poloidally, and {\it vice versa}.

\section{Conclusions}
By evaluating the unity, $\cos$, and $\sin$ flux surface moments of the particle and force density equations, we develop a system of analysis which elucidates the radial and poloidal profiles of various quantities in a steady state tokamak plasma discharge.  Applying our analysis to a particular shot, we discover evidence that plasma viscosity is not well modeled by a Newtonian fluid approach.  Nonlinear pseudoplastic behavior is observed, indicating regions of comparatively thin and glassy plasma may coexist in a single discharge, and between species.  Glassy plasma may explain many phenomena, such as the formation of transport barriers in high performance discharges, as well as the structure of the edge pedestal.  Further verification of our present analysis by analyzing a helium-carbon plasma discharge with fully corrected velocity measurements is suggested.

\begin{acknowledgments}
The author gratefully acknowledges the helpful comments of W.~M.~Solomon, and the contributions of the DIII-D Team who made the measurements presented in this paper.  This work is supported by the GTF Endowment.
\end{acknowledgments}

\appendix*
\section{Force Density Equations}\label{app:eqns}
\subsection{Toroidal Unity Moment}
\begin{align} \label{eqn:torunity}
 & n_j^0 m_j \left[ V_{\vphi \, j}^0 \left( \frac{1}{r} V_{r \, j}^0 + \frac{\partial}{\partial r} V_{r \, j}^0 \right) + V_{r \, j}^0 \frac{\partial}{\partial r} V_{\vphi \, j}^0 \right] + m_j V_{\vphi \, j}^0 V_{r \, j}^0 \frac{\partial}{\partial r} n_j^0 & \\
 & + n_j^0 m_j \sum_{k \neq j} \nu_{jk}^0 \left(V_{\vphi \, j}^0-V_{\vphi \, k}^0 \right) + \frac{n_j^0 \Tbar_j V_{\vphi \, j}^0}{\nu_{\vphi \, j}^0 R_0^2}
- n_j^0 e_j \left(B_\vtheta^0 V_{r \, j}^0+E_\vphi^0\right)
- M_{\vphi \, j}^0
= 0 & 
\end{align}

\subsection{Toroidal Cosine Moment}
\begin{align} \label{eqn:torcos}
 & n_j^0 m_j \sum_{k \neq j} \nu_{jk}^0 \left[\left(2+\wt{n}^c_k+\wt{n}^c_j+\wt{V}_{\vphi \, j}^c\right) V_{\vphi \, j}^0-\left(2+\wt{V}_{\vphi \, k}^c+\wt{n}^c_k+\wt{n}^c_j\right) V_{\vphi \, k}^0\right] & \\
 & + m_j \left(\wt{V}_{\vphi \, j}^c+2+\wt{n}^c_j\right) \left(n_j^0 V_{r \, j}^0 \frac{\partial}{\partial r}V_{\vphi \, j}^0+V_{\vphi \, j}^0 V_{r \, j}^0 \frac{\partial}{\partial r}n_j^0+V_{\vphi \, j}^0 n_j^0 \frac{\partial}{\partial r}V_{r \, j}^0\right) & \\
 & + \frac{n_j^0 m_j V_{\vphi \, j}^0}{r} \left[\left(\wt{V}_{\vphi \, j}^c+4+\wt{n}^c_j\right) V_{r \, j}^0+\left(\wt{V}_{\vtheta \, j}^s+\wt{V}_{\vphi \, j}^s+\wt{n}^s_j\right) V_{\vtheta \, j}^0\right] & \\
 & - \frac{n_j^0 \Tbar_j V_{\vphi \, j}^0}{\nu_{\vphi \, j}^0 R_0^2} \left[21\wt{n}^c_j \wt{V}_{\vphi \, j}^s/4 +\wt{V}_{\vphi \, j}^s\left(3+1/\epsi^2\right) +3\wt{n}^s_j\left(1-\wt{V}_{\vphi \, j}^c\right)/4 \right]  & \\
 & -  n_j^0 e_j \left[\left(1+\wt{n}^c_j\right) B_\vtheta^0 V_{r \, j}^0+\left(\wt{n}^c_j+2\right) E_\vphi^0 \right]
- 2 M_{\vphi \, j}^0
= 0 & 
\end{align}

\subsection{Toroidal Sine Moment}
\begin{align} \label{eqn:torsin}
 & n_j^0 m_j \sum_{k \neq j} \nu_{jk}^0 \left[\left(\wt{n}^s_j+\wt{n}^s_k+\wt{V}_{\vphi \, j}^s\right) V_{\vphi \, j}^0-\left(\wt{n}^s_j+\wt{V}_{\vphi \, k}^s+\wt{n}^s_k\right) V_{\vphi \, k}^0 \right] & \\
 & + m_j \left(\wt{n}^s_j+\wt{V}_{\vphi \, j}^s\right) \left(V_{\vphi \, j}^0 V_{r \, j}^0 \frac{\partial}{\partial r}n_j^0+V_{\vphi \, j}^0 n_j^0 \frac{\partial}{\partial r}V_{r \, j}^0+n_j^0 V_{r \, j}^0 \frac{\partial}{\partial r}V_{\vphi \, j}^0\right) & \\
 & + \frac{n_j^0 m_j V_{\vphi \, j}^0}{r} \left[\left(\wt{n}^s_j+\wt{V}_{\vphi \, j}^s\right) V_{r \, j}^0-\left(2+\wt{V}_{\vtheta \, j}^c+\wt{V}_{\vphi \, j}^c+\wt{n}^c_j\right) V_{\vtheta \, j}^0 \right] & \\
 & + \frac{n_j^0 \Tbar_j V_{\vphi \, j}^0}{\nu_{\vphi \, j}^0 R_0^2} \left[ \left(3\wt{n}^c_j/4 -1/\epsi^2\right)\left(\wt{V}_{\vphi \, j}^c-1\right)-15\wt{V}_{\vphi \, j}^s \wt{n}^s_j/4 \right]  & \\
 & - n_j^0 e_j \wt{n}^s_j \left(B_\vtheta^0 V_{r \, j}^0+E_\vphi^0\right)
= 0 &
\end{align}

\subsection{Poloidal Unity Moment}
\begin{align} \label{eqn:polunity}
 & m_j \left(n_j^0 V_{r \, j}^0 \frac{\partial}{\partial r}V_{\vtheta \, j}^0+V_{\vtheta \, j}^0 n_j^0 \frac{\partial}{\partial r}V_{r \, j}^0+V_{\vtheta \, j}^0 V_{r \, j}^0 \frac{\partial}{\partial r}n_j^0\right) + 2 \frac{n_j^0 m_j V_{\vtheta \, j}^0 V_{r \, j}^0}{r} & \\
 & + \frac{n_j^0 \Tbar_j \tau^\star_j}{B_\vphi^0 R_0^2} \left[ \frac{1}{3}\left(2-\wt{V}_{\vtheta \, j}^c \right) V_{\vtheta \, j}^0 B_\vphi^0+\left(-1+\wt{V}_{\vphi \, j}^c\right) V_{\vphi \, j}^0 B_\vtheta^0 \right] + \frac{n_j^0 \Tbar_j}{2 R_0} \epsi \wt{n}^s_j & \\
 & + n_j^0 m_j \sum_{k \neq j} \nu_{jk}^0 \left(V_{\vtheta \, j}^0-V_{\vtheta \, k}^0\right) 
+ n_j^0 e_j  2 B_\vphi^0 V_{r \, j}^0
= 0 & 
\end{align}

\subsection{Poloidal Cosine Moment}
\begin{align} \label{eqn:polcos}
 & n_j^0 m_j \sum_{k \neq j} \nu_{jk}^0 \left[ \left(1+\wt{n}^c_j+\wt{V}_{\vtheta \, j}^c+\wt{n}^c_k\right) V_{\vtheta \, j}^0-\left(1+\wt{n}^c_k+\wt{n}^c_j+\wt{V}_{\vtheta \, k}^c\right) V_{\vtheta \, k}^0 \right] & \\
 & + m_j \left(1+\wt{n}^c_j+\wt{V}_{\vtheta \, j}^c\right) \left(V_{\vtheta \, j}^0 V_{r \, j}^0 \frac{\partial}{\partial r}n_j^0+n_j^0 V_{r \, j}^0 \frac{\partial}{\partial r}V_{\vtheta \, j}^0+V_{\vtheta \, j}^0 n_j^0 \frac{\partial}{\partial r}V_{r \, j}^0 \right) & \\
 & + \frac{n_j^0 m_j V_{\vtheta \, j}^0}{r} \left(2 V_{r \, j}^0 \wt{V}_{\vtheta \, j}^c+V_{\vtheta \, j}^0 \wt{n}^s_j+3 V_{r \, j}^0+2 V_{\vtheta \, j}^0 \wt{V}_{\vtheta \, j}^s+2 \wt{n}^c_j V_{r \, j}^0\right) & \\
 & + \frac{n_j^0 \Tbar_j \tau^\star_j}{R_0^2} [ ( 7 \wt{n}^s_j \wt{V}_{\vtheta \, j}^s/12+4 \wt{V}_{\vtheta \, j}^c/3-4/3-5 \wt{n}^c_j \wt{V}_{\vtheta \, j}^c/4+4 \wt{n}^c_j/3 ) V_{\vtheta \, j}^0 & \\
 & + ( 2-\wt{n}^s_j \wt{V}_{\vphi \, j}^s + 2 \wt{n}^c_j \wt{V}_{\vphi \, j}^c-2 \wt{n}^c_j-2 \wt{V}_{\vphi \, j}^c ) V_{\vphi \, j}^0 B_\vtheta^0/B_\vphi^0 & \\
 & + ( \frac{1}{3} ( \wt{V}_{\vtheta \, j}^c-2 ) V_{\vtheta \, j}^0+ ( 1-\wt{V}_{\vphi \, j}^c ) V_{\vphi \, j}^0 B_\vtheta^0/B_\vphi^0 ) /\epsi^2 ] & \\
 & + \frac{n_j^0 \Tbar_j}{r} \wt{n}^s_j
+ n_j^0 e_j B_\vphi^0 V_{r \, j}^0 \wt{n}^c_j
= 0 &
\end{align}

\subsection{Poloidal Sine Moment}
\begin{align} \label{eqn:polsin}
 & n_j^0 m_j \sum_{k \neq j} \nu_{jk}^0 \left[ \left(\wt{V}_{\vtheta \, j}^s+\wt{n}^s_k+\wt{n}^s_j\right) V_{\vtheta \, j}^0-\left(\wt{V}_{\vtheta \, k}^s+\wt{n}^s_k+\wt{n}^s_j\right) V_{\vtheta \, k}^0 \right] & \\
 & + m_j \left(\wt{V}_{\vtheta \, j}^s+\wt{n}^s_j\right) \left(V_{\vtheta \, j}^0 n_j^0 \frac{\partial}{\partial r}V_{r \, j}^0+V_{\vtheta \, j}^0 V_{r \, j}^0 \frac{\partial}{\partial r}n_j^0+n_j^0 V_{r \, j}^0 \frac{\partial}{\partial r}V_{\vtheta \, j}^0\right) & \\
 & - \frac{n_j^0 m_j}{r} \left[ (V_{\vtheta \, j}^0)^2-(V_{\vphi \, j}^0)^2-2 V_{\vtheta \, j}^0 V_{r \, j}^0 (\wt{n}^s_j+\wt{V}_{\vtheta \, j}^s)+2 (V_{\vtheta \, j}^0)^2 \wt{V}_{\vtheta \, j}^c+(V_{\vtheta \, j}^0)^2 \wt{n}^c_j \right] & \\
 & + \frac{n_j^0 \Tbar_j \tau^\star_j}{R_0^2} [ (\wt{V}_{\vtheta \, j}^s/3 \epsi^2+7 \wt{n}^s_j \wt{V}_{\vtheta \, j}^c/12+19 \wt{n}^c_j \wt{V}_{\vtheta \, j}^s/12) V_{\vtheta \, j}^0 & \\
 & - (3 \wt{n}^c_j \wt{V}_{\vphi \, j}^s+\wt{V}_{\vphi \, j}^s/\epsi^2) V_{\vphi \, j}^0 B_\vtheta^0/B_\vphi^0 ] & \\
 & - \frac{n_j^0 \Tbar_j}{r} \wt{n}^c_j
+ n_j^0 e_j  B_\vphi^0 V_{r \, j}^0 \wt{n}^s_j
= 0 &
\end{align}

\newpage



%

\newpage

\begin{enumerate}
\item Figure 1. Closed path line integral of $E_r$ and $E_\vtheta$.
\item Figure 2. Input profiles for poloidal magnetic field, toroidal electric field, toroidal momentum injection, C6 toroidal velocity, C6 poloidal velocity, temperature, and density.
\item Figure 3. Radial electric field profiles for both deuterium and carbon.
\item Figure 4. Poloidal asymmetry coefficients for density, poloidal velocity, and toroidal velocity, for the friction dominated solution.
\item Figure 5. Toroidal and poloidal rotation and viscosity profiles for the friction dominated solution.  The effective viscosities are labeled ``drag'', and the theoretical viscosities are labeled ``gyro'' and ``para''.
\item Figure 6. Poloidal asymmetry coefficients for density, poloidal velocity, and toroidal velocity, for the viscosity dominated solution.
\item Figure 7. Toroidal and poloidal rotation and viscosity profiles for the viscosity dominated solution.  The effective viscosities are labeled ``drag'', and the theoretical viscosities are labeled ``gyro'' and ``para''.
\item Figure 8. The functional dependence of $\tau_j^{\star}$ on $V_{\vtheta \, j}$ for (a) deuterium and (b) carbon.
\end{enumerate}

\clearpage
\newpage

\begin{figure}%
\includegraphics{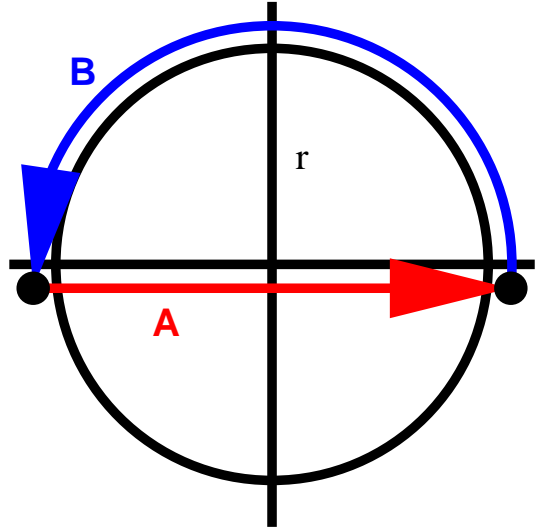}
\caption[Closed path line integral of $E_r$ and $E_\vtheta$.]{\label{fig:circle}(Color online). Closed path line integral of $E_r$ and $E_\vtheta$.}
\end{figure}

\begin{figure}%
\includegraphics{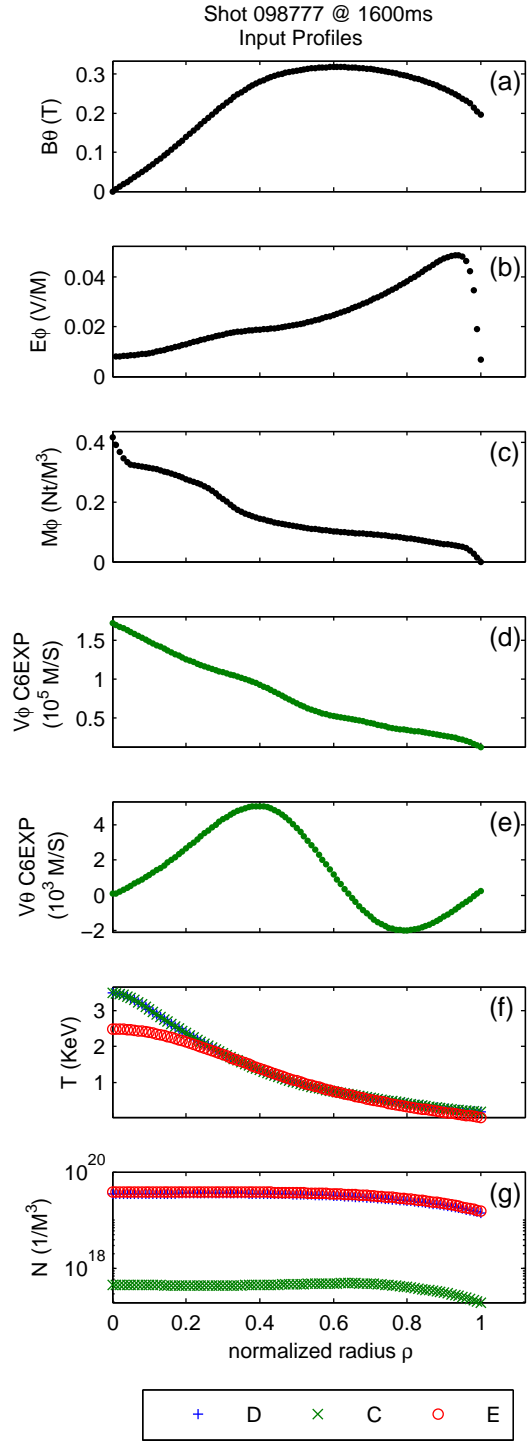}
\caption[Input profiles.]{\label{fig:input}(Color online).  Input profiles for poloidal magnetic field, toroidal electric field, toroidal momentum injection, C6 toroidal velocity, C6 poloidal velocity, temperature, and density.}
\end{figure}

\begin{figure}%
\includegraphics{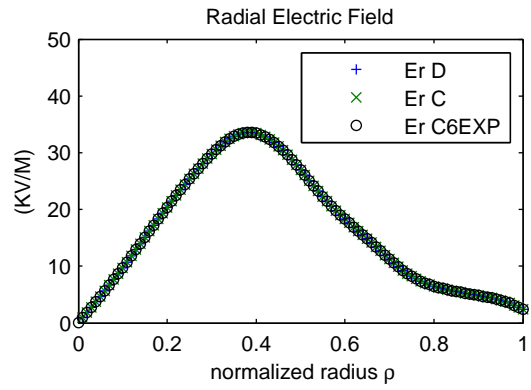}
\caption[Radial electric field profiles for both deuterium and carbon.]{\label{fig:erad}(Color online).  Radial electric field profiles for both deuterium and carbon.}
\end{figure}

\clearpage

\begin{figure*}%
\includegraphics{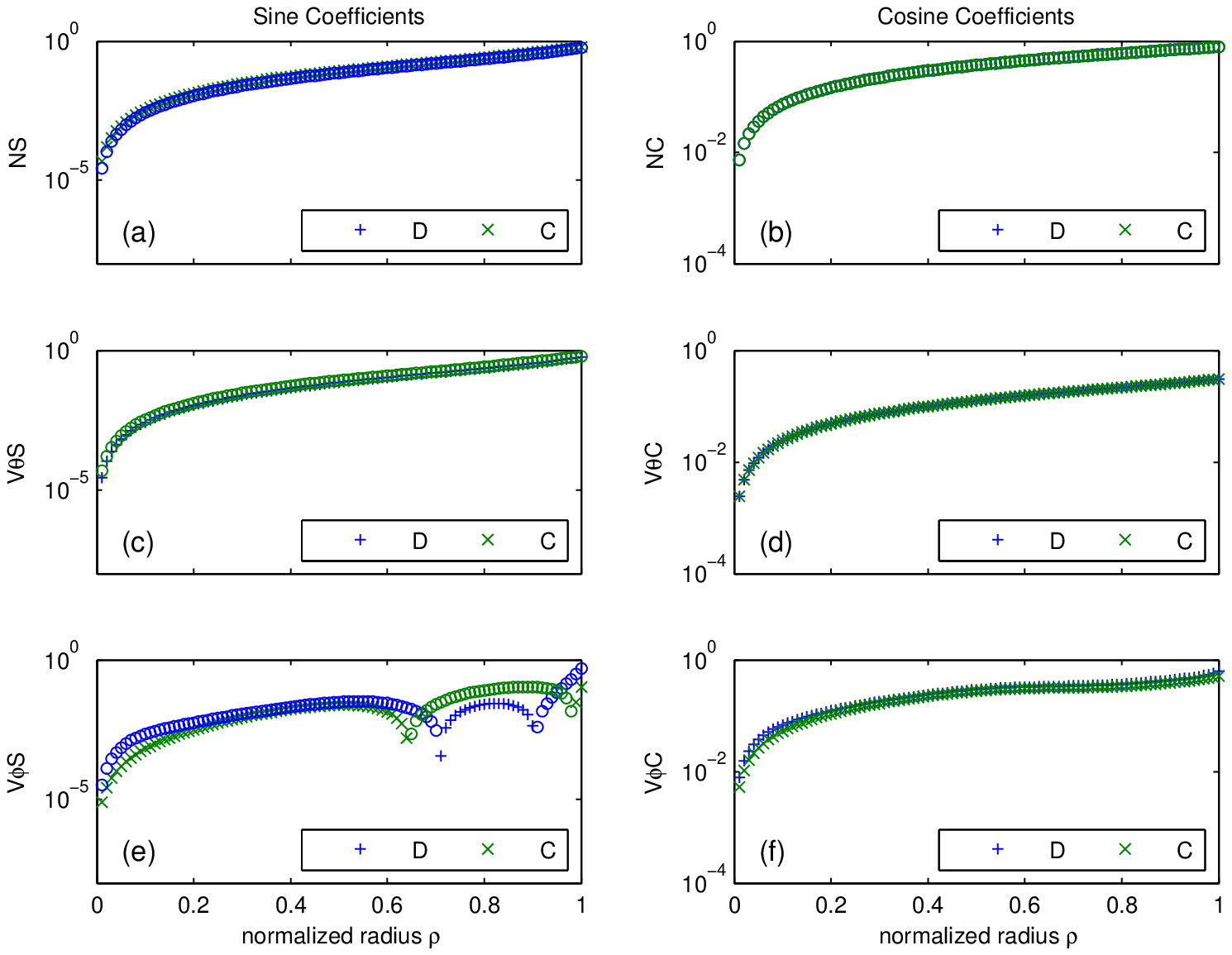}
\caption[Poloidal asymmetry coefficients for the friction dominated solution.]{\label{fig:polfric}(Color online).  Poloidal asymmetry coefficients for density, poloidal velocity, and toroidal velocity, for the friction dominated solution.}
\end{figure*}

\begin{figure*}%
\includegraphics{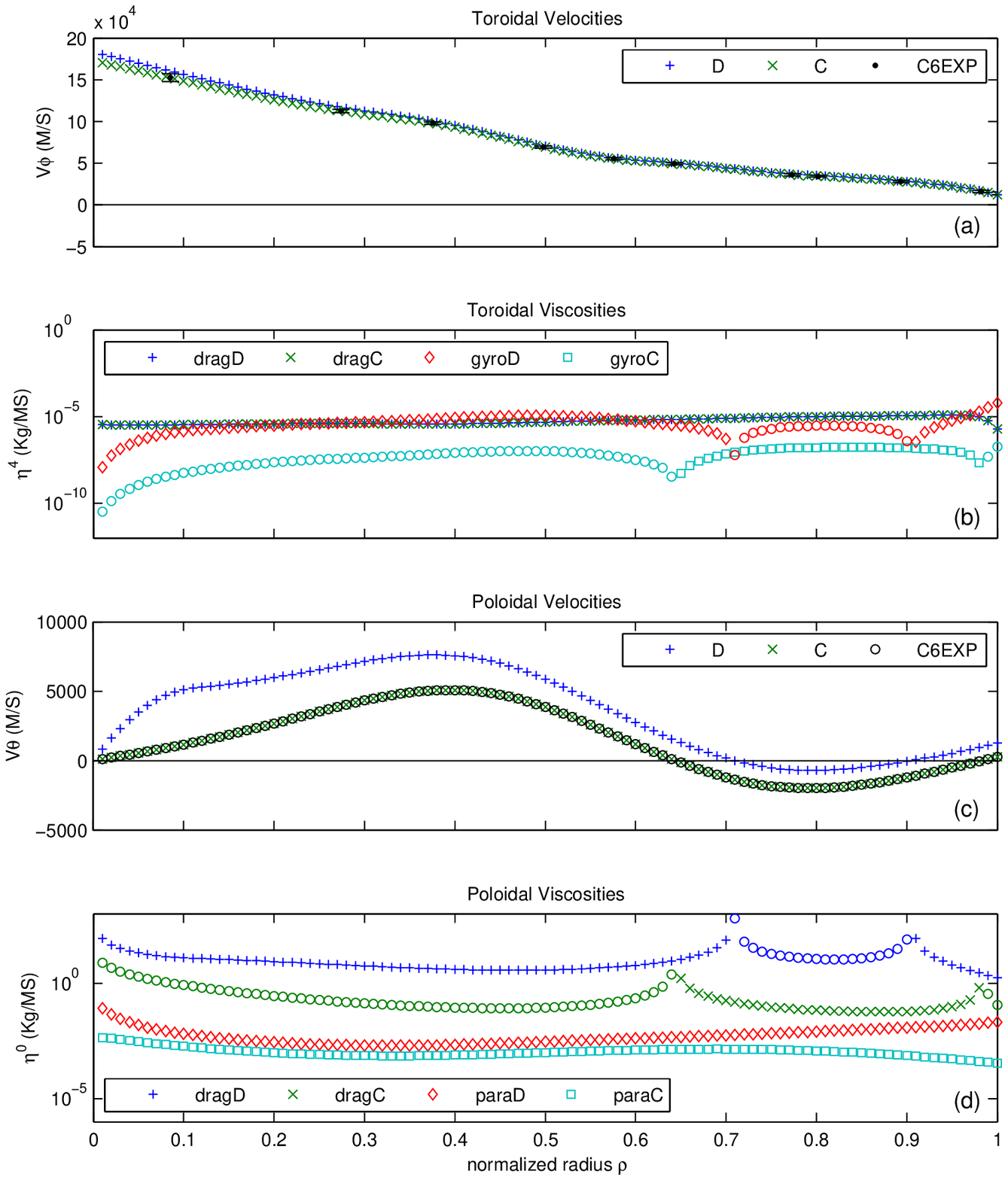}
\caption[Rotation and viscosity profiles for the friction dominated solution.]{\label{fig:velfric}(Color online).  Toroidal and poloidal rotation and viscosity profiles for the friction dominated solution.  The effective viscosities are labeled ``drag'', and the theoretical viscosities are labeled ``gyro'' and ``para''.}
\end{figure*}

\begin{figure*}%
\includegraphics{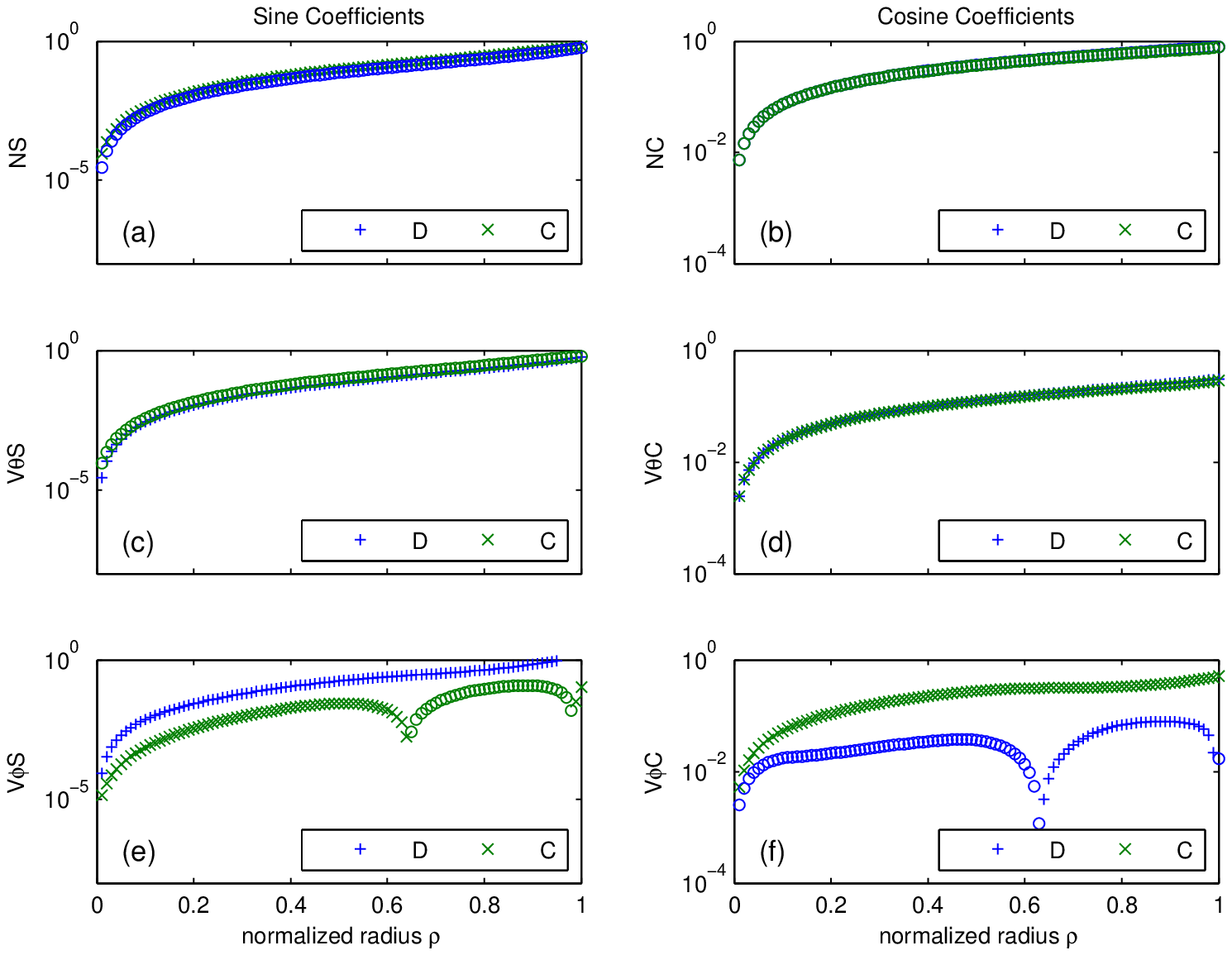}
\caption[Poloidal asymmetry coefficients for the viscosity dominated solution.]{\label{fig:polvisc}(Color online).  Poloidal asymmetry coefficients for density, poloidal velocity, and toroidal velocity, for the viscosity dominated solution.}
\end{figure*}

\begin{figure*}%
\includegraphics{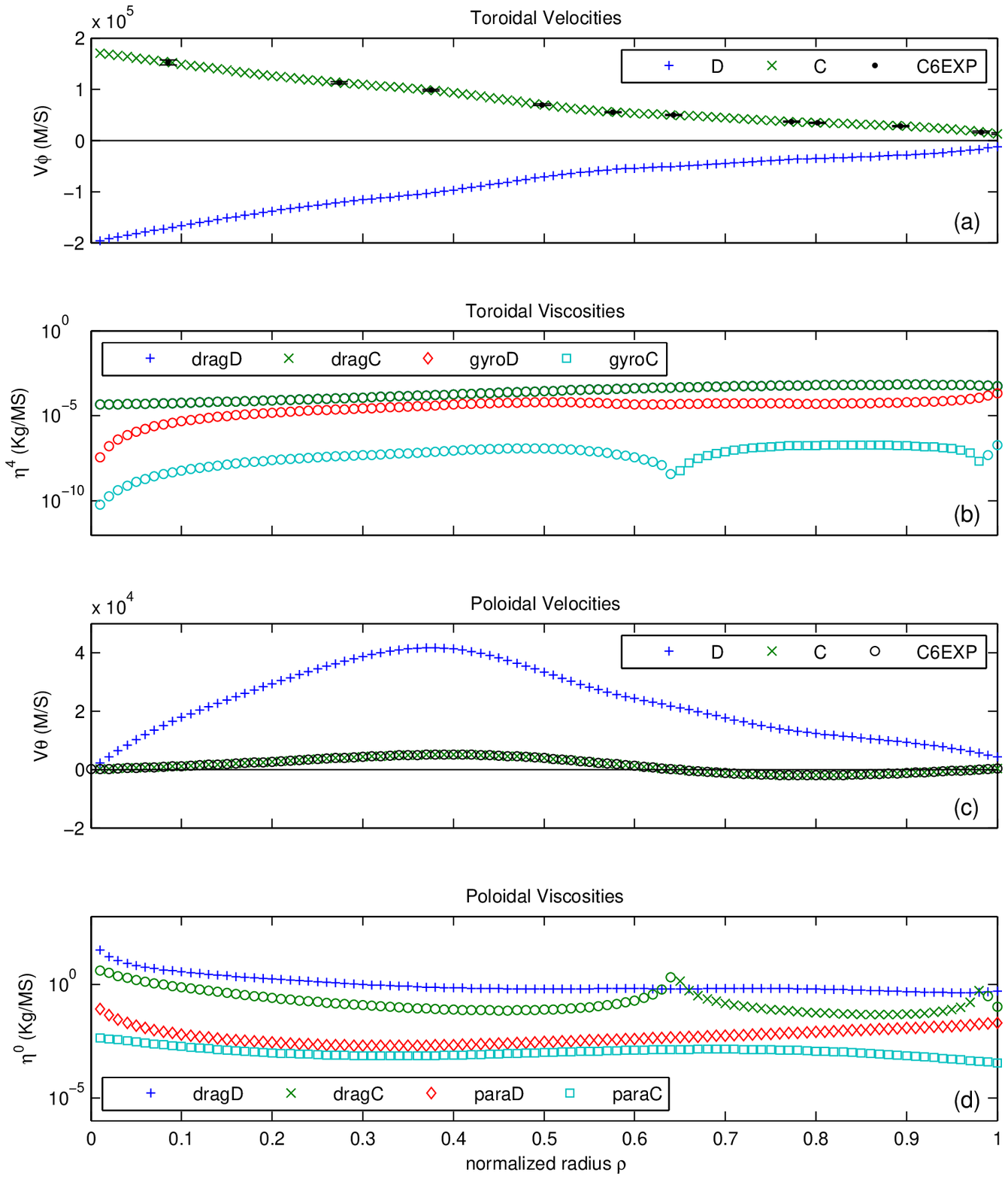}
\caption[Rotation and viscosity profiles for the viscosity dominated solution.]{\label{fig:velvisc}(Color online).  Toroidal and poloidal rotation and viscosity profiles for the viscosity dominated solution.  The effective viscosities are labeled ``drag'', and the theoretical viscosities are labeled ``gyro'' and ``para''.}
\end{figure*}

\begin{figure*}%
\includegraphics{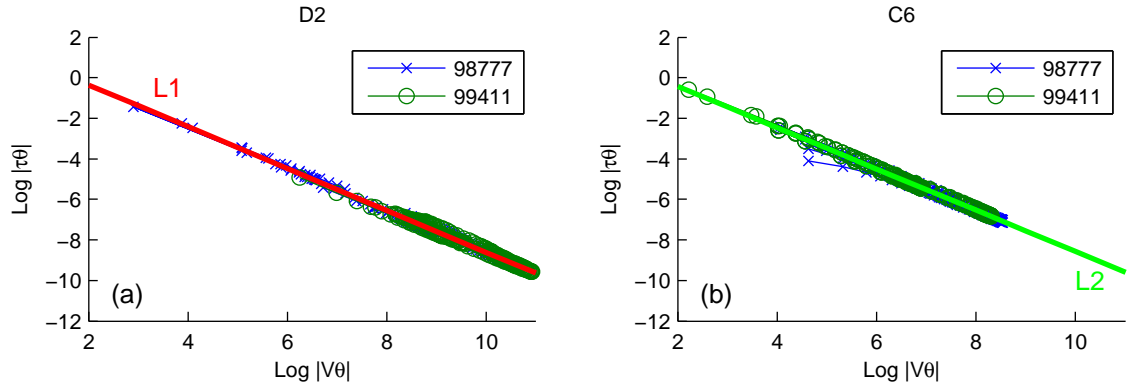}
\caption[The functional dependence of $\tau_j^{star}$ on $V_{\vtheta \, j}$.]{\label{fig:tauvsV}(Color online).  The functional dependence of $\tau_j^{\star}$ on $V_{\vtheta \, j}$ for (a) deuterium and (b) carbon.}
\end{figure*}

\end{document}